\documentclass[reqno]{amsart}
\usepackage{amssymb,amsmath,amsfonts,amsthm,graphics,ifthen,mathrsfs}
\usepackage[mathcal]{euscript}
\usepackage[cmtip,arrow]{xy}
\usepackage{graphicx}
\usepackage{epsfig}
\usepackage{psfrag}

\begin{document}

\title{Probabilistic Methods on Erdos Problems}
\author{Jesse D. Gilbert}
\date{5 November, 2026}

\maketitle

\begin{abstract}
Probability distributions and the rules of inclusion-exclusion apply to derive general results about graph theory and number theory. In order to apply the rules of inclusion-exclusion in the setting some of these problems require, it is necessary to use specific equations and make several assumptions about integration and measure of the probability distributions. \\

\noindent \textbf{Keywords:} Probability Distribution, Differential Equation, Measure \\
\noindent \textbf{AMS Classification:} 05C78
\end{abstract}

\section{Main Result}

\noindent Make a general counting argument that there is a labeling $\phi$ from a set of labelings $\phi \in \Phi$ such that $\phi(\{H_i\})$ has no duplicate labels on the fibers $e$ of $\{H_i\}.$ Then the random variable $X$ that measures duplications of labels on the fibers has $0<P[X=d=0].$ Let $2\mu$ be the expected number of duplicate labels. Let $\alpha$ be the number of fibers on $\{H_i\}.$ Finally, let $d \alpha$ be the cardinality of the comparison orbit on the $\{H_i\}.$ The cardinality of the comparison orbit is either a function of the the strength of the labeling $\phi,$ or some parameter that measures the scope of fibers that each fiber is compared to when the duplicate labels are measured. Define the measure of the expectation over each fiber, $E[X_i],$ as $\beta = (C \cdot  \mu)/(\nu_0)$ where $\nu_0=\alpha/(d \alpha),$ and the constant of proportionality $C$ is usually either $2$ or $4,$ depending on the labeling problem, and the type of fibers that are selected, which are usually either vertices or edges. Construct a product topology on the fibers $e$ of $\{H_i\}$ equal to $[X_i = d \cdot d \alpha]^{\times}.$ The goal is to show that there is an element of this product topology where every argument is $0,$ and that this corresponds to a 'free' labeling $\phi \in \Phi,$ a labeling that does not register any label duplications over the labeling $\phi(\{H_i\}).$ In order for all the arguments of the vectors in the product topology to measure label duplications in the same coloring (using the same $s$ colors), the value $d$ has to be multiplied by $d \alpha,$ otherwise the probability measured on the vectors has units $(\textnormal{(labelings per)}~d \alpha)^{\alpha}.$ See that these last types of labelings do not correspond to labelings with $s$ labels. The value of $C$ is an example of a natural correspondence, and cannot be inflated, in order to make the proof technique easier to use. Consider the sums of the random variables from the product space and write the sums as $\sum_{[t_0]} X_i.$ Refer to $\sum_{[t]}X_i=f(X)$ when $|[t_0]|$ is $|[t]|.$ Outline a proof technique that generates a lower bound on the percentage of free labelings from $\Phi$ using $s$ labels. Check that $P[f(X_i)=0]$ has $e^{-1/\beta}/\beta  \le P[f(X_i)=0],$ for all $i,$ and write this condition as $e^{-1/\beta}/\beta  \le P[X_1=0],$ or just $1/\beta << P[X_1=0]$ in most cases. Notice that it is necessary to check this basis condition on the random variables for each of the individual fibers \textit{before} doing any identification of the fibers in the product space. An inclusion-exclusion argument will establish that $$1/|S| \le \Gamma(|[t]|) \cdot P[\sum_{[t]}X_i=0],$$ and the conclusion that $0<P[X=0]$ will follow under some general set of conditions. Start with $f_1(x) \circ \omega \le g_1(x),$ where $g_1$ is defined at the measure $\omega.$ Then, apply Equation (\ref{E:EA}) to the underlying distributions:

\begin{equation} \label{E:EA}
\sum_{i \in [t]} \frac{d}{dk} f_{t,i} (x)=\sum_{[t]-i \subset [t]}\sum_{j \in [t]-i} f_{t-1,j}(x)- \sum_{i \in [t]} f_{t,i}(x).
\end{equation}

\noindent Here, $f_{t,i}(0)$ is the number of free labelings on $[t]$ of the $\{X_i\}.$ Each value $f_{t,i}$ is based at some element $i,$ and $f_{t,i}=f_{t,j}$ when the values $f_{t,i}$ and $f_{t,j}$ are taken from the same set of $[t]$ $X_i$, and $i,j \in [t].$ The set $[t]$ on $f_{t,i}$ should always be clear from context. From the solutions of Equation (\ref{E:E1}), it follows that if $e^{-1/\beta}/\beta \le f_{1,i}(0)$ for all $i,$ then $e^{-1/\beta}/\beta^{|[t_0]|} \le f_{t_0,i}(0)$ for all $i \in [t_0],$ and for all $t_0 \le t,$ and $[t_0] \subset [t].$ Start with Equation (\ref{E:EA}). Take $\min_{\{[t-1]\}}~\{f_{t-1,i}(x)\}=f_{t-1}(x)$ where the subset $[t-1] \subset [t]$ is arbitrary, and $i \in [t-1]$ varies. Then use $f_{t,i}=f_{t,j}$ and solve for $f_{t,i}(x).$ There is some $f_t(x)$ that satisfies both $f_t(x) \le f_{t,i}(x),$ and Equation (\ref{E:E1}) on the support $[0,b).$ Apply 

\begin{equation} \label{E:E1}
 \frac{d}{dk} f_t(x)=|[t-1]|f_{t-1}(x)-f_t(x)
\end{equation}

\noindent where $k=x/\beta$ to some $f_{t}(x) \le f_{t,i}(x).$ It follows that if $e^{-1/\beta}/\beta \le P[X_1=0]$ on all the fibers of induction, then $e^{-1/\beta}/\beta^{|[t]|} \le \Gamma(|[t]|) \cdot P[\sum_{[t]}~X_i=0].$ It remains to find solutions to Equation (\ref{E:E1}). Use the principle of inclusion-exclusion:

\begin{equation} \label{E:EB}
R_{[t]}(x)=|[t-1]| S_{[t-1]}(x)-S_{[t]}(x).
\end{equation}

\noindent Let $Y_j = \sum_{[t]-j} X_i$ and suppose $\bigcap$ is taken over $j \in [t] \setminus [1].$ Then

\begin{equation} \label{E:EF1}
S_{[t]}(x)= \Gamma(|[t]|) \cdot P[\sum_{[t]} X_i \le x \bigcap Y_i \le x],
\end{equation}

\noindent and 

\begin{equation} \label{E:EF2}
R_{[t]}(x)= \beta \cdot \Gamma(|[t]|) \cdot P[x< \sum_{[t]} X_i \bigcap Y_i \le x].
\end{equation}

\noindent To see Equation (\ref{E:EF2}), notice that the right-hand side of Equation (\ref{E:EB}) has the union over the sample sets underlying the $S_{[t-1]}(x).$ Count the elements of the sample set represented by $P[\sum_{[t]}~X_i=x_0]$ $x_0$ times. The overcount of the underlying sample set of $R_{[t]}(x)$ relative to the sample set associated with $S_{[t]}(x)$ is $\beta:$ the integral over $x P[\sum_{[t]}~X_i = x]$ with respect to $x,$ divided by $|\{X_i\}|.$ Try to show

\begin{equation} \label{E:E2}
\frac{d}{dk} S_{[t]}(x)=R_{[t]}(x).
\end{equation}

\noindent To derive results for a mixed distribution that satisfies Equation (\ref{E:EB}), use a sequence of functions (whose derivatives are dirac-delta functions) that converge point-wise to the mixed distribution. Apply Equations (\ref{E:EA}) and (\ref{E:E1}), according to the differential $k=x/\beta;$ the limiting behavior is given by applying a measure directly to Equation (\ref{E:E1}). Let $$x_1=P[x < \sum_{[t]}X_i \bigcap Y_i \le x]=x_{1,1}+x_{1,2}+x_{1,3}$$ where

$$x_{1,1}=P[x< \sum_{[t]}X_i <cx \bigcap Y_i \le x]$$
$$x_{1,2}=P[cx \le  \sum_{[t]}X_i  \le cx \bigcap Y_i \le x]$$
$$x_{1,3}=P[cx < \sum_{[t]}X_i \bigcap Y_i \le x].$$

\noindent There is an additional partition. Use the two partitions to make the equality clear, and there is the use of limits. Let $$x_2=P[x+\epsilon < \sum_{[t]} X_i \bigcap Y_i \le x+ \epsilon]=x_{2,1}+x_{2,2}+x_{2,3}$$ where

$$x_{2,1}=P[x+ \epsilon <  \sum_{[t]} X_i <c(x+ \epsilon) \bigcap Y_i \le x+\epsilon]$$
$$x_{2,2}=P[c(x+ \epsilon)  \le  \sum_{[t]} X_i  \le c(x+ \epsilon) \bigcap Y_i \le x + \epsilon]$$
$$x_{2,3}=P[c(x+ \epsilon) < \sum_{[t]} X_i \bigcap Y_i \le x+\epsilon].$$

\noindent Then take the limit in the following expression: $$\lim_{\epsilon \rightarrow 0} \frac{1}{\epsilon} [(x_{2,1}+ x_{2,2}+ x_{2,3})-(x_{1,1} +  x_{1,2} + x_{1,3})].$$ 

\noindent Then the limit $$\lim_{\epsilon \rightarrow 0} \lim_{c \rightarrow 1} \frac{1}{\epsilon}[(x_{2,1}- x_{1,1})]$$ has both terms such that the argument of the probability reduces to the null set as $c$ goes to $1.$ So both parts of this terms are identically zero because the argument of the probabilities are equal to the null set. This is the easiest reduction in the expression that begins the calculation. Some other reductions are summarized next. \\

\noindent Take the limit of $c$ to the value 1 in $$\lim_{\epsilon \rightarrow 0} \lim_{c \rightarrow 1} \frac{1}{\epsilon}[(x_{2,2}- x_{1,2})].$$ See that $(d/dk)(d/dx) S_{[t]}(x)$ is precisely equal to $(d/dx) R_{[t]}(x):$ \\

\noindent It is necessary to expand the two terms in the limit expression $$\lim_{\epsilon \rightarrow 0} \lim_{c \rightarrow 1} \frac{1}{\epsilon}[(x_{2,3} - x_{1,3})]$$ to see that this term is equal to zero. \\

\noindent Write $$x_{2,3}=x_{2,3,a}+x_{2,3,b}$$ and write $$x_{1,3}=x_{1,3,a}+x_{1,3,b}.$$ Then set the equalities 

$$x_{1,3,a}=P[cx < \sum_{[t]} X_i \le c(x+\epsilon) \bigcap Y_i \le x]$$
$$x_{1,3,b}=P[c(x+\epsilon) < \sum_{[t]} X_i \bigcap Y_i \le x].$$ Set the equalities 

$$x_{2,3,a}=P[c(x+\epsilon) < \sum_{[t]} X_i \bigcap x < Y_i \le x+\epsilon]$$
$$x_{2,3,b}=P[c(x+\epsilon) < \sum_{[t]} X_i \bigcap Y_i \le x].$$ See that the difference between $x_{2,3,b}$ and $x_{1,3,b}$ is zero, and see that as $\epsilon$ goes to zero in $x_{1,3,a}$ and $x_{2,3,a},$ one of the conditions in each of the probabilities is null in the limit. Then the limit $$\lim_{\epsilon \rightarrow 0} \lim_{c \rightarrow 1} \frac{1}{\epsilon}[(x_{2,3} - x_{1,3})] \longrightarrow 0$$ (regardless of the factor $1/\epsilon$ in front of the expression) goes to zero. \\

\noindent Then it follows that $$ \frac{d}{dk} S_{[t_0]}(x)=R_{[t_0]}(x).$$ See this (picture proof) by drawing a picture of two annuli and associating the limit expression as the difference in radius between the two individual bands that separate the depth of each annulus into two regions. Then the preceding argument is just a formalization of this picture proof. \\

\noindent The solution to Equation (\ref{E:E1}) using $(d/dx)P[\sum_{[t]} X_i \le x]=P[\sum_{[t]} X_i =x]$ gives the following expression: The solution depends on Equation (\ref{E:EF2}), and the form of the second derivative of the cumulative distribution function underlying $S_{[t]}(x).$ Then

\begin{equation}\label{E:E3}
f_t(x)=x^{|[t-1]|}e^{-x/\beta}/\beta^{|[t]|}.
\end{equation}

\noindent This assumes that $f_1(x)=e^{-x/\beta}/\beta$ at every point of the support $[0,\infty).$ The Equation (\ref{E:E3}) gives a scalar multiple times a probability distribution function, and the integral of this product solves Equation (\ref{E:E1}), given assumptions about the form of the function $f_1(x)$. Generally, functions $g_t$ such that $f_1(x) \le g_1(x)$ at every point of the support $[0,b)$ where the $\{g_t\}$ obey Equation (\ref{E:E1}), have $f_t(x) \le g_t(x)$ for  $x \in [0,b)$ and all $1<t.$ Use reduction ad absurdum. \\

\noindent Furthermore, for some $\{g_t\}$ that satisfy Equation (\ref{E:EB}), defined on particular measures $\omega$ considered here, it holds that if $f_1(x) \circ \omega \le g_1(x)$ on some support $x \in [0,b),$ then it follows that $f_t(x) \circ \omega \le g_t (x)$ on that same support $x \in [0,b).$ Apply the measure $\omega$ to the Equation (\ref{E:E1}). It may be that $[(d/dk) f_t(x)] \circ \omega \not = (d/d(\omega/\beta))[f_t(x) \circ \omega].$ \\

\noindent In order for there to be a first $x \in [0,b)$ where $g_t (x) < f_t \omega(x),$ both the value $h_t \omega(x)=g_t (x)-f_t \omega(x) < 0,$ and $[(d/dk)h_t(x)] \circ \omega < 0.$ These inequalities imply that $h_{t-1} \omega (x)=g_{t-1} (x)-f_{t-1} \omega (x) < 0,$ which is a contradiction to inductive assumption. Use Equations (\ref{E:EA}) and (\ref{E:E1}) and the inequality $f_{t-1}\omega(x) \le g_{t-1} (x),$ to get $f_{t} \omega(x) \le g_{t} (x),$ and $g_{t_0} (x) \le \min_{\{[t_0]\}} \{\Gamma(|[t_0]|) \cdot P[\sum_{[t_0]}~X_i \le x] \}.$ \\

\noindent No probability density function will obey equality with $(d/d\omega) f_t \omega$ at all points of the support, because $1<1/[\beta(1-e^{1/\beta})].$ \\

\noindent Apply the following steps to show $$1/|S| \le g_t(x)$$ where $g_t(x)$ is $\Gamma(|[t]|)$ times the cumulative distribution function of $\sum_{[t]}X_i$ at the value $x.$ Go on to explain the components of the proof, including some explanation of the portions that have to be carried out on an individual basis.  \\

\noindent (Step 1.) Show there is a 'uniform map' from $V(H) \longrightarrow V(K_{n(H)})$ that is uniform on the fibers of $\{H_i\}.$ This requires a variety of conditions, and it is necessary to verify that the product topology applies as described. \\

\noindent (Step 2.) Calculate $\beta \varpropto \mu/\nu_0$ where $\nu_0=\alpha/(d \alpha)$ and $\mu=E[X(d)].$ Usually, $\beta$ is proportional by some small multiple of $2$ that is appropriate to the counting in various settings of labeling problems. \\

\noindent (Step 3. Part One.) Verify that $g_1$ has $$\sum_{x \le x_0} \frac{d}{d\omega} f_1(x) \circ \omega \le g_1(x_0)$$ on some support $x \in [0,x_0].$ This will invoke induction. \\

\noindent (Step 3. Part Two.) This has already been verified: Equation (\ref{E:E1}) applies to $\{f_t\}$ and functions underlying the $\{g_t\}.$ Observe that, for instance, $$e^{-1/\beta}/\beta^{|[t]|} \le g_t(0)$$ as a consequence of the first two steps and the first part of the third step. \\

\noindent (Step 3. Part Three.) Verify that $$1/|S|< g_{t}(0)$$ where $|S|=|\Phi|.$ \\

\noindent (Conclusion) Verifying each of the individual steps in the reference method demonstrates that there is some $\phi \in \Phi$ that labels the elements of the domain of the $\{\phi\}$ so that there are no fibers in the original graphs $\{H_i\}$ that register label duplications. It is necessary to elaborate on some of the complications. Verify the steps on a case by case basis. The main assumption required is that if the probability of an event with less than $x$ intersections is greater than the reciprocal of the cardinality of the probability space, then the event necessarily occurs with less than $x$ intersections. There is a natural correspondence between the factorial of the fibers, and a factor in the denominator of the cumulative distribution function $P[\sum_{[t]} X_i \le x].$ The antecedent of the main assumption, the reference set hypothesis, is tested as $$\frac{1}{|S|}\frac{1}{\Gamma(\alpha)} < P[\sum_{[t]} X_i \le 0].$$ The function $f_t'(x)$ is either continuous, or is, rather, based on a mixed distribution that is integrated against a measure different from $dx.$ In either case, the cardinality of the solution set of $P[\sum_{[t]} X_i \le x]$ is $\Gamma(\alpha)$ times the cardinality of the solution set of $f_t(x).$ \\

\noindent Because of the natural correspondence between the factorial of the fibers and the factor (gamma function) in the denominator of the cumulative distribution, the test for the reference set hypothesis usually simplifies to $1/|S| \le e^{-1/\beta}/\beta^{|[t]|}.$ See that if $P[\sum_{[t]} X_i \le 0]=P[\sum_{[t]} X_i < 1],$ then $$\frac{1}{|S|}\frac{1}{\Gamma(\alpha)}<\frac{1}{\Gamma(\alpha)} \frac{e^{-1/\beta}}{\beta^{|[t]|}}$$ is the test for the reference set hypothesis. As a result of this natural correspondence between the solution set of $P[\sum_{[t]} X_i \le x]$ and the factorial of the fibers, it follows that, after checking Step 1 and Step 2, the only conditions required are in Step 3 of the reference set method: 

\begin{equation} \label{E:E5}
e^{-1/\beta}/\beta \le P[X_1=0]
\end{equation} 

\noindent and

\begin{equation} \label{E:E6}
1/|S| \le e^{-1/\beta}/\beta^{|[t]|}
\end{equation}

\noindent (in the case that $P[\sum_{[t]} X_i \le 0]=P[\sum_{[t]} X_i <1]$). \\

\noindent To derive the continuous function $f_t(x),$ assume that $(d/dx)S_{[t]}(x)=(d/dx)f_t(x)$ and write that 

\begin{equation} \label{E:E7} 
S_{[t]}(x)=x^{|[t-1]|}e^{-x/\beta}/\beta^{|[t]|}.
\end{equation}

\noindent A consequence is that $S_{[t]}(x)=f_t(x).$ \\

\noindent The duplication units are the domain of the random variable that counts duplications on the labeled graph. The number of duplications at the duplication units are collected at the fibers of the graph (or model). The fibers and the duplication units are usually chosen from either the edges or vertices. The cardinality of the comparison orbit is not always well-defined for every phrasing of every problem. In order to do the inclusion-exclusion argument, the fibers have to be identified, and in order for the fibers to be properly identified, the elements of the domain of $\phi$ have to induce the fibers of induction one by one in some resolution of the graph or model. Otherwise, the problem has to be rephrased. \\

\noindent In the case that $\phi$ labels the vertices of $G$, and the duplication units are edges, the problem can be phrased so that the fibers of induction are vertices. In this case, every duplication at every duplication unit has two images, one for each endpoint of each edge. Then when $P[X_1=0]$ is counted at the vertices, usually the entire edge neighborhood of each vertex has to be free of duplications. In these phrasings, the value $P[X_1=0]$ has to account for duplications at both the endpoints of each edge, or two images of each duplication at each edge. When $\phi$ is on $V(G)$ and the duplication units and fibers are at $E(G),$ and the graph is acyclic, then the elements of the domain of $\phi$ induce the fibers of induction one by one. In these cases, the problem can be rephrased so that the labeling is on the edges of $G.$ If $G$ is not acyclic, then this auxiliary construction will not work, because the cardinality of the comparison orbit of one or more fibers has to be taken equal to infinity. However the problem is phrased, it follows that when the comparison orbit is well-defined, and the fibers of induction can be properly identified, then the problem is said to be well-defined. The calculations here are consistent with the notion that the cumulative distribution function is continuous on some of the support. The cumulative distribution function can take on values that are not equal to $r/|S_0|$ where $r$ is an integer, and $|S_0|$ is the cardinality of the sample set. Interpret this to mean that there are $r$ elements of the sample set that satisfy $X \le x$ when $x$ is the minimum value such that $r/|S_0| \le P[X \le x].$ 

\section{Examples}

\noindent Work out several examples, so the reader can see how the reference set method is applied to specific situations. In some cases, write $X_1=r,$ $P[X_1=r=x_0],$ and $E[X_1=r]=\beta$ to denote a change in the definition of the random variable. \\

\noindent (Example. Estimating Brooks' Theorem) Try deriving a close approximation to Brooks' Theorem. \\

\noindent Step 1. The variable is phrased as a local variable on vertex fibers. Vertices of degree $\Delta(G)$ produce the worst case for the basis step. \\

\noindent Step 2. The value for $P[X_1=0]$ is usually much larger than $$1/(\Delta(G)+1)=1-\frac{\Delta(G)}{\Delta(G)+1}.$$ (Put $s=\Delta(G)+1.$) Additionally, the value of $\beta$ should be $2 \mu/(n/d \alpha).$ That is, the relevant expectations for the problem are $$\mu=\frac{n}{2} \cdot \frac{\Delta(G)}{\Delta(G)+1}$$ and $\beta \approx \Delta(G)$ since $d \alpha=\Delta(G)+1.$ Because the vertex label duplications occur in pairs, the factor of 2 in the calculation of beta from $\mu$ gives this last expression. In general, when $\Delta(G)< s,$ the condition will be satisfied. Set $s=\Delta(G)+1.$ \\

\noindent Step 3. It should generally follow that $e^{-1/\beta}\beta^{-n}$ is greater than $(\Delta(G)+1)^{-n}$ for the values of $\beta$. The second obstacle is the basis step. In the case described here, $e^{-1/\beta}/\beta \le 1-E[X_1] \le P[X_1=0]$ where $E[X_1]=\Delta(G)/[\Delta(G)+1].$ The first inequality in the pair of inequalities should be fairly close. 

\newpage

\noindent (Example. Weakened Bollobas-Eldridge Conjecture) Verify the steps, and check the result. \\

\noindent Step 1. Take $\pi(\phi(\{H_i\})) \prec (n-1)^n.$ There are only two graphs; the problem is a packing problem.  Take $V(K_n)$ to $V(K_n)$ uniformly, and take the vertices as the fibers of induction. The random variable has comparison orbit $d \alpha = n:$ The value of $d \alpha$ in $\nu_0=\alpha/(d \alpha)$ is $|V(K_n)|.$ There is, for this problem, an improvement available, but do not use it in this setting. \\

\noindent Step 2. Take $\Delta(G)\Delta(H)<< n-2,$ and the standard formulation for $\beta:$ Use $\beta=4 \mu/\nu_0$ where $\nu_0=n/n.$ The number of fibers is $n$. The parameter $\mu$ is given by the Sauer and Spencer expectation formula for packing graphs, and is a global parameter. Obtain $$\mu \sim (1/2) \Delta(G)\Delta(H)\frac{n}{(n-1)} + O(1);$$ set $\beta=n/e$. Then $2 \cdot e[\Delta(G)\Delta(H)]n/(n-1) \le n.$\\

\noindent Step 3. To check the basis step, consider the number of duplications $X$ at a vertex of the placement of $G$ and $H.$ Furthermore take, $$1 - \frac{\Delta(G)\Delta(H)}{(n-1)}< P[X_1=0].$$ Then the relevant calculation is $$1/n << (1-\frac{\Delta(G)\Delta(H)}{(n-1)})$$ for $\Delta(G) \Delta(H) << n-2.$ Then to check the reference set hypothesis, choose $\beta$ much smaller than $n$ because on the left-hand side of the check for the reference set hypothesis, there is an $n!$ in the denominator and not an $n^n.$ \\

\noindent The conclusion, $2 \cdot e [\Delta(G)\Delta(H)] + O(1) \sim n$ as sufficiency for packing $G$ and $H$ in $K_n$, is weaker than the Bollobas-Eldridge Conjecture. The conclusions are also weaker than the criteria for packing $G$ and $H$ in $K_n,$ due to Sauer-Spencer, $2 \Delta(G) \Delta(H) \le n$; consider the check on the hypothesis. The problem considered here highlights an important limitation of the reference set method, when it is applied to graph labeling problems that are graph packing problems. In this case, the vertices were taken as fibers, and generally there is no limitation in that case. However, if the graph labeling problem is a graph packing problem with multiple elements in the packing, and the edges are taken as fibers, then the problem generally needs to be rephrased so that there is an additional degree sequence requirement. These types of problems are examined in the example here, but the vertices were taken as the fibers, so that there is no degree sequence requirement. Generally, when the edges are taken as fibers, the graphs have to have a packing as a multigraph whose degree sequence is majorized by the sequence $(n-1)^n,$ in order for the reference set method to give any useful results. In addition to this special consideration, there are also modifications to the general method that have applications in graph packing problems when the packing problems have only two elements. These special types of packing problems have been studied extensively, and the Sauer-Spencer criteria is one of the outstanding results. 

\newpage

\noindent The problem of calculating the expectation is not always very obvious to resolve. There is generally some natural correspondence between $\mu$ and $\beta,$ but this correspondence can vary with the context of the labeling problem. The next two examples use a formalization of the calculation for the expectation. It is helpful to motivate the calculation of the expectation in a particular way to standardize calculations, and make the verification of calculations easier to perform. The concept of an inventory for calculating expectation has a direct and concrete application in both of the two examples that follow directly. The concept also has applications in the $\rho$-labeling and $h$-labeling problems included here as well. While the inventory is not mentioned explicitly in these two other examples that follow, the expectations in these other two examples were calculated using the inventory $I=\{2K_2,P_3\}.$ \\

\noindent Go on to look at various complete graph selection problems: Try the following expression for $\mu:$ $$\mu=\sum_{G \in I} E_1[G,H] \cdot E_2[G,H].$$ Then, roughly, $E_1[G,H]$ is the number of labeled copies of $G \subseteq H,$ and $E_2[G,H]$ is the expected number of edge-duplicate pairs taken over the number of labeled copies of $G$ in $H.$ Then $E_1[G,H]=|L(G) \subset H|.$ Define $$E_2[G,H]=\frac{|\{(e,e'):\phi(e)=\phi(e'),~e \not =e',~\phi(G) \subset H\}|}{|\{\phi(G) \subset H\}|}.$$  

\noindent Then in the case that $\mu$ is on $H \cong K_n$ and $I=\{2K_2,P_3\},$ use the equality $$\mu=n_4/8 \cdot E_2[2K_2,K_n]+n_3/2 \cdot E_2[P_3,K_n].$$ It is not clear why an inventory of graphs with more than 2 edges would be required. In this case, if $\mu$ is taken over 2-edge pairs, it would have to be that no two graphs in the inventory shared the same 2-edge subgraph. Generally, $I=\{2K_2,P_3\}.$ This idea for $\mu$ is meant to be completely general. It is difficult to accomplish the desired over-arching generality. Try to be systematic, and give a process that is verifiably correct. In some instances, this goal is accomplished. The ideas are from Burnside-Frobenius counting, also called Polya counting. The results on the Golomb ruler problem and Ramsey theory problem outlined here are strictly partial results. The results here serve to verify, motivate, and add context to the general study of each problem. In particular, the Ramsey problem is more flexible, but the ideas here are not well-developed enough to answer bichromatic Ramsey problems for graphs in general. It might be interesting to try to address the problem of finding $R(P_n,P_n)$ using these techniques or some other species of graph or tree, but there are no obvious applications from the results here. Calculating the cardinality of the reference set would differ dramatically from problem to problem in the context of Ramsey theory, and because the vertices would presumably be taken as the fibers, there would have to be some system for mapping the duplications from the duplication units of edges to the vertices. Not only do the phrasings of vertex labeling problems have different correspondences when the duplication units are edges, and the fibers are either vertices or edges, but some polychromatic labeling problems have other allowances that are not easy to explain. After looking at the next two examples, the reader may want to try looking at the polychromatic diagonal Ramsey problem. These other types of Ramsey problems have complications that are not easy to explain.

\newpage

\noindent (Example. The Golomb Ruler Problem) The three steps follow. This example shows that the calculations in one of the steps can break down in some cases, affecting the applicability of the steps to the problem as a whole. Charted examples are taken from Wikipedia, and help show how the method applies to Golomb rulers. Set $s=G(n).$ See evidence that $s \sim n^2+O(n^{3/2}),$ asymptotically \cite{Dollas}.\\

\noindent Step 1. It suffices to consider that the symmetric map from $K_n$ to $K_n$ is uniform on the vertices of $K_n.$ This map will carry the edges of $K_n$ to the edges of $K_n$ uniformly. In this case, the target graph and the graph being selected are isomorphic, making it clear that the symmetric map will generate a uniform map without any trouble. \\

\noindent Step 2. Use $\beta=4 \mu/\nu_0$, because the duplication units are edges and the fibers are vertices. To calculate the constant of proportionality, there is an additional factor of two where the edges intersect the fibers at two points. Using the expectation formula, count $$\mu= n_4/8 \cdot E_2[2K_2,K_n]+ n_3/2 \cdot E_2[P_3,K_n] \approx \frac{1/6 \cdot n_4 + 1/8 \cdot n_3}{s}.$$ Use integration to make it easier to approximate the expectations involved. To calculate the expectation, it helps to use integration in both the $P_3$ and $2K_2$ step. The reader can reconstruct the counting techniques used to count the expectations by observing the completed integral in the first step. That is, to calculate $E_2[2K_2,K_n],$ in this case, use $$E_2[2K_2,K_n] \approx \frac{4 s_2}{s_4 \cdot s^2} \cdot \int_0^{s} x^2~dx.$$ Then, approximate $E_2[2K_2,K_n] \approx 4/(3 s).$ Use a similar approach to estimate $E_2[P_3,K_n]$ in this case: Count $E_2[P_3,K_n] \approx 1/(4 s).$ Estimate $s$ and $\beta$ by assuming that $s \sim \beta +O(n).$ Now take $s_0 \cdot \beta \sim 2/3 (n^4 - 5.25n^3+O(n^2))$. Using the formula for $\mu$, the formula for $\beta,$ and using $n$ as the number of fibers of $K_n$ (use the vertex-set) and algebra, $\beta \approx \sqrt{2/3}(n^2-2.625n+O(n)).$ There is a root 2, since $s$ and $\beta$ are supposed asymptotic, and so the coefficient of the second term is divided by $2.$\\

\noindent Step 3. Show that under some conditions, $s \approx \sqrt{2/3}n^2.$ Take $$P \approx \frac{ s(s-1)(s - (2+1)) \dots (s - {n \choose 2}) }{[s(s-1)(s-2) \dots (s-n+1)]}.$$ Try estimating $P \approx (s-X)^n/(s_n)$ where $X \approx (1/6)[n^2].$ Since the value subtracted from $s$ in the numerator of $P$ is large, this should be a significant over-estimate. The probability $P$ counts the probability that a set of $n$ edges, all sharing a common vertex, so that they form a star, taken from the difference basis $[s],$ do not have any edge label duplicates among the star edges either between the star edges or between the star edges and the pendant clique suspended from the end-vertices of the star. There could be edge label duplicates among the clique edges in the clique suspended on the end-vertices of the star. \\

\noindent Derive the value of $P$ by counting the number of available labels for pendant vertices on a labeled star. The labeling necessarily avoids edge difference duplications between any of the first $t_0$ labeled pendant edges and any other edge differences generated in the clique on the first $t_0$ labels. Assume there were no duplicates among the first few ($t_0-1$) labels. The derived estimate of $s=G(n) \approx \sqrt{2/3}(n^2-2.625n)$ results from the included calculations. See charted values: the derived estimate is not always an upper bound for experimentally determined values when $3 \le n \le 27.$ The experimental values for $s$ do match the derived values closely in this range. It is not clear if the calculation here supports the determined experimental value $G(21)=333.$ One model based on the reference set method predicts that $G(21) \le 320,$ using the correct $n^2$ and $n$ terms in the expansion for $\mu;$ the model may generate an artificially small value for $s,$ because the value for $E_2[2K_2, K_n]$ is only an estimate. Computer searches determined the experimental value of $G(21),$ and it is not clear if the experimentally determined $G(21)$ is purported to be an exact value, as opposed to an upper bound. See \cite{WKPD1}, \cite{Distributed}. There are cases where the trial value of $s=G(n)$ is not sufficiently large to verify the basis step. In these cases, increase the value of $s.$ The value for $\beta$ will actually decrease, raising the threshold for $P[X_1=0].$ However, the estimate for $P[X_1=0]$ increases as well. The value $s =  \lceil \sqrt{2/3}(n^2-2.625n)   \rceil$ is not sufficient to pass the basis case for $n=22.$ Try amending the general term to to $(s-(25/24){n \choose 2})$ as opposed to $(s-{n \choose 2}).$ Then the calculation for $P$ more accurately reflects the value of $P$ for small orders. Assume the ordering of the first $(n-1)$ choices is arbitrary, and assume that the $n^{th}$ choice falls arbitrarily in the order according to a uniform distribution. \\

\noindent Count the expected number of edge label duplications on the difference between the $x_i$ (inserted) label and the $x_j$ label and differences between $x_h$ and $x_k$ labels $(\ge \max \{x_i, x_j\})$ or the $x_h$ and $x_k$ labels $(\le \min \{x_i, x_j\})$. This is counted by a complicated sum on the order of $n,$ which is estimated by an integral. \\

\noindent That is, count the remaining number of expected duplications on the label between $x_i$ and $x_j$, where $x_i$ is the $n^{th}$ label in the order, and $x_j$ ranges on the order. Each of the new $(n-1)$ differences can be duplicated once by the pre-existing differences, since those pre-existing differences are all distinct. The count on the differences between the chords all generated by the new point is $(n-1)(n-2)/2={(n-1) \choose 2}.$ Take the expression $$E[X] \approx (n-1)+{(n-1) \choose 2} + \frac{2\int_0^n \int_0^h h^2/2+(h-x)^2/2 ~dx ~dh}{(\int_0^n dh) (\int_0^n dh)},$$ as a rough estimate of the number of differences that have to be avoided every time a label is selected. Then $E[X] \approx {n \choose 2}+n^2/48 \approx (25/24){n \choose 2}.$ The average choice is about $(s-(25/24){x \choose 2}).$ A consequence of changing the model in this fashion is that the theoretical model and the current experimental model share exactly one value, $G(21)=333.$ Modifying the model more accurately reflects the value $G(21)$ and accepts a basic discrepancy in the uniform distribution on the vertex labels. Again, from \cite{Dollas}, it appears that basic calculations can establish that $G(n) \sim n^2 +o(n^{1+\epsilon})$ is an asymptotic on $G(n)$ for large $n.$ Upon changing the model to incorporate this different value for $P,$ the relative error between the two curves, for small values, increases from a value around three or four percent for large values, to a value around nine or ten percent. \\

\newpage

\noindent (Example. Diagonal Ramsey Theory) Outline the three steps in the proof. The example helps to refine the understanding of the method overall, but there are some points of contention. First, the formula for $\mu$ is used. Take $\beta=4 \mu/(\alpha/d \alpha)$ where $\mu$ is defined by the counting formula in $E_1$ and $E_2$ already outlined and $\alpha/d \alpha =1.$ (In other cases than the bichromatic case, the whole scheme needs to be reconsidered because of the polychromatic counting.) The vertices are taken as the fibers. This problem can be phrased so that the edge-duplicate pairs are accounted for by multiplying the number of pairs by two and then, to account for the endpoints, multiplying the pairs by two again because of dimensional analysis. Again, in the higher polychromatic cases, the counting is different at a number of points, and it is not absolutely clear how to reconcile these cases. Take $\alpha=n.$ \\

\noindent Step 1. Take the vertices of $K_n$ to $K_n$ by the symmetric group $S_n.$ The map is uniform on the vertices and edges of $K_n.$ \\

\noindent Step 2. Use $\beta=4 \mu/\nu_0$ and count $\mu$ in a very specific way. Use $K_3$ in the inventory and then count only duplications in two of the overlapping $P_3$ in the $K_3$ because this is sufficient to indicate bichromaticity in complete graphs. The formula for $\mu$ is $$\mu=n_3/6 \cdot E_2[G,H].$$ However, take $E_2[G,H]=1$ instead of the $3/2$ indicated by counting independent distributions of $1/2$ under the binomial expansion in three edges. Map one copy of a $P_3$ to each of two of the respectives endpoints of the triangle. If there is an edge label mismatch in the triangle, then one of the two $P_3$ rooted at these opposing endpoints will indicate the label mismatch. Choosing an edge of the triangle, and the endpoints of the edge as the targets of the mapping, map the copies $P_3 \subset K_3$ so that the group $C_n$ acts on the correspondences of the triangles in the collection of triangles as transitively as possible. Take for granted that this is possible on the collection of triangles. Then at each of the two endpoints of the triangle, there is an image of one of the $P_3$ contained in the triangle and $E_2=1$ indicates the probability that there is an edge label mismatch in the $P_3.$ Then $\beta=(n_3/6) \cdot 1 \cdot 4=2n_3/3$ where at each of the endpoints there are four units that count each of the possible mismatches, two for each edge of the mismatch. Thus $\beta$ counts 4 or 8 for each triangle with a mismatch. The expectation $\beta$ counts some triangles with mismatches more than others, but counts at least 4 for each triangle with a mismatch. Then try to use these counting units in Step 3 to verify $e^{-1/\beta}/\beta \le P[X_1=0].$ \\

\noindent Step 3. Using the units ($P_3 \subset K_3$) from Step 2, the probability that the units rooted at a vertex are monochromatic is $2^{(n-1)_2/3},$ assuming independence of the edge label probabilities on the $P_3$ in the $K_3$, a $1-1$ ratio of labels, and the assumption that if each unit is monochromatic, and if the first unit is monochromatic, then all the units are monochromatic of the same color or label. This predicts $P[X_1=0]<1/\beta$ for the first time at $n=6$ (and that $101.5937... \le \beta \not = s$). It is not exactly clear if it is necessary to use the same units for $X_1$ as the expectation. Alternatively, it might appear that the calculation would break down at $(2/3)n_3 < 2^{n-2};$ this is not consistent, according to the vertex degrees of $K_n.$ The proximity of this predicted value of $s$ for $n=6$ is a good feature, but it might be misleading for a number of reasons. Then if $\lceil \beta \rceil =102,$ it should be $s \le \lceil \beta \rceil +3=105$ for $s=K(6,6).$

\newpage

\noindent (Example. $\rho$-Labeling Problems) Try deriving approximations to the $\rho$-labeling problem. \\

\noindent Step 1. Set up the problem for the edge-difference labeling of trees, or graphs more generally, using the $\rho$-labeling scheme. Note that there is a restriction on the graphs that are considered here, so that the ideas in this example are limited to trees and forests. Take the edges to a copy of $K_{2s+1}$ uniformly. Then the map carries the edges of $H$ to the edges of $K_{2s+1},$ and use the product topology on the edges of the graph $H.$ In this case, the modular structure of the packing that results is not utilized. Rather, establish a product topology on the edges of $H.$ The map is uniform on the edges and vertices of $H.$ \\

\noindent Step 2. Take the value for $\mu$ as $$\mu = 1/2 \sum_{e} \frac{m-1-d(e)}{s-1/2}+ 1/2 \sum_{e} \frac{d(e)}{2s-1}.$$ Use the edges of the graph as the fibers, and assume there is one fixed graph under the labeling scheme. Then the sum is given by the summation of the expected values over the graphs in the inventory and here the probability of a duplication is related to the number of allowable labels; take $[2s+1]$ to be the labeling set. Instead of taking $d \alpha = 2s,$ it appears valid to take $d \alpha =s,$ because there are two copies of each comparison class in the difference classes at each vertex. This calculates the allowance on the cardinality of the comparison orbit $d \alpha$. Then $\nu_0=m/s,$ and $\beta=2\mu/\nu_0.$ Set $s$ equal to $m$ so that the cardinality of the comparison orbit is identical to the cardinality of the number of fibers. The value $\beta << m-1$ for sufficiently large $n.$ The cardinality of the set of fibers of induction is $m=|[t]|=\alpha.$ \\

\noindent Step 3. Under these conditions, it is not possible to immediately conclude that an acyclic $(n,n-1)$-graph has a $\rho$-labeling. (The acyclic $(n,n-1)$-graphs are just the trees.) When the basis step is verified, see that to preserve generality, $d(e)$ has to be taken $d(e)=1,$ because the basis step has to be performed before the identification of the fibers. The base case does not check. There are cases where $$e^{-1/\beta}/\beta << 1-(m-3/2)/(s-1/2) \le P[X_1=0]$$ does not necessarily hold. \\

\noindent Notice that it is also not sufficient to extend the support from $[0,1]$ to $[0,e],$ and then try to show the integral of the test function on the support $[0,e]$ is less than $P[X_1=0].$ The tolerance of $f(X)=\sum_{[t]} X_i$ will give $P[f(X) \le t \cdot e],$ using the support in this fashion and $t \sim n.$ Here, this is not an important point because $\beta$ is sufficiently smaller than $2s+1.$ That is, the reference set hypothesis checks because $\beta<<2s+1.$ See that it is possible to choose $E[d(e)]$ so that it is very large relative to $m.$ Because of the form of $\mu$, which has different denominators on the edge-degree term for the $2P_2$ term and the $P_3$ term, choosing $E[d(e)]$ larger gives a smaller choice of $\beta$. For instance, take trees with small radius. The implication may be that, if automorphism is ignored, then trees with large choices for $E[d(e)]$ have more $\rho$-labelings. Compare the star to the path. No correspondence between radius and absolute number of labelings has been demonstrated conclusively. \\

\newpage

\noindent (Example. $h$-Labeling Problems) Try deriving approximations to the popular harmonious labeling problem for trees. \\

\noindent Step 1. Define a near-harmonious labeling as a map from the vertex set of $H$ to $[n(H)]$ such that the sum at each edge modulo $n$ is distinct at each edge. If $H$ is a tree, there will be a missing edge-sum. Form a harmonious labeling from any near-harmonious labeling by rotating the zero label to any vertex of the tree. Add to the vertex labels modulo $n$, and then add the missing edge and edge sum to the desired vertex. This construction gives the form of a harmonious labeling. So, try to show any tree has a near-harmonious labeling. Establish a uniform map by rotating $H$ around $K_{n}.$ \\

\noindent Step 2. Take $s=n(H).$ Again, in a fashion analogous to the previous example, take the worst-case value for $\mu$ as $$\mu \sim 1/2\sum_{e} \frac{m-1-d(e)}{s-3}+ O(1).$$ There is no summand for the $P_3$-term because the additive labeling scheme here cannot have duplicates at $P_3$s. Let $\nu_0=m(H)/s$, set the fibers of induction equal to the set of edges, and consider the cardinality of the comparison orbit equal to the number of modular additive classes at each vertex. Not every modulus is available at every vertex in the simple graph version of $K_n.$ The sum for $\mu$ is given by the summation of the expected number of edge label duplications over $2K_2$ and $P_3$.\\

\noindent Step 3. The value of $\beta$ appears dependent only asymptotically on $s$. The base case does not appear to hold in all cases, since $m \le s-1,$ and $1 \le d(e)$ are sharp. When $s$ is actually taken close to $n,$ and $E[d(e)]$ is small, there are cases where it holds that $\beta^{-m}<1/s_{(m+1)},$ so that the reference set hypothesis will not check in all generality. The hypotheses of the reference set method require the graph be a forest. \\

\noindent Consider the permutations on the fibers generated by considering the inductive orbits and inductive sub-orbits of the fibers. Order the resolutions of the graph so that they are each trees, and then consider the fibers as differences between the resolutions. When the graph under resolution is acyclic, the comparison orbit is well-defined, and the fibers can be properly identified. That is, for the $\rho$-labeling problem, and the $h$-labeling problem, it is possible to show the comparison orbit is well-defined for a forest. \\

\noindent When $\phi$ is on $V(G)$ and the edges are taken as the fibers, some of the important elements of the theory break down for non-forests. Show that the requirement $e^{-1/\beta}/\beta \le P[X_1=0],$ is fulfilled, and then it is generally required that differences of consecutive resolutions generate the individual fibers. When a stage of the resolution induces multiple chords (instead of a single edge), such as when there is any cycle in the graph, the general conditions on the comparison orbit fail, and this can cause problems with the counting argument. \\

\noindent The argument that follows is consistent with the assumption that the two variables at a fiber, $X_1$ and $X_2,$ are not independent.

\newpage

\noindent (Example. Vizing's Theorem) Apply the method to Vizing's Theorem. \\

\noindent Step 1. When $G$ is not regular, construct a super-graph $G'$ of $G$ using extra nodes and edges, so that the super-graph is regular of degree $\Delta(G).$ Map the auxiliary graph to a subgraph of $K_{n(G')}$ using the symmetric map. The symmetric map from the auxiliary graph to itself is uniform, because $G'$ is regular. After completing the steps, try to conclude that $G'$ obeys Vizing's Theorem. It will follow that the original graph $G$ obeys Vizing's Theorem. \\ 

 \noindent Step 2. Use $\Delta(G)=\Delta,$ fix $d \alpha=s,$ and fix $s=\Delta+1.$ Then fix $E[X_1]$ on the graph equal to the expected number of edge label duplicates at a node of $G.$ Then $E[X_1]=(\Delta-1)/(\Delta+1).$ Set $$2 \mu=\frac{n \Delta}{2} \cdot \frac{(\Delta-1)}{(\Delta+1)}.$$ 

\noindent Step 3. The value for $\beta \approx \Delta$ will follow in this Step 3, and depends on $E[X_1]$ across both nodes of an edge. For $E[X_1]$ at a node, it appears that $(1-2P[X_1+X_2=0]) \le E[X_1],$ whenever $X_1$ and $X_2$ are the random variables at either endpoint of an edge. In the case $s = \Delta(G)+1,$ it follows that $1/(\Delta(G)+1) \le P[X_1+X_2=0].$ It appears valid to take $d \alpha = s$ in the given instance. Use the Boolean variable at an edge so that $E[X_1+X_2] \le x/(\Delta(G)+1).$ The basis case is satisfied under the last inequality. (Consider various values of $x.$) The reference set hypothesis holds, and there is a labeling with no edge label duplications when $s$ is set to $s=\Delta+1.$ That is, the base case holds, and $1/|S|=1/(\Delta+1)^{\alpha} < e^{-1/\beta}/\beta^{\alpha}.$ \\

\noindent The general result is the strongest possible of this type. Odd cycles require $\Delta(G)+1$ labels for a proper edge labeling. The traditional proof of Vizing's Theorem is constructive, and is very difficult. The inequality $(1-2P[X_1+X_2=0]) \le E[X_1]$ is roughly equivalent to taking $E[X_1]^{1/2}$ at a node for the upper bound on the disjunction of two probabilities. This point motivates the inequality, but there appear to be examples where $1-E[X_1]^{1/2} \not \le P[X_1+X_2=0].$ Rather, $(1-2P[X_1+X_2=0]) \le E[X_1]$ is verified strictly by considering the value $E[X_1].$ \\

\noindent Observe that the random variables $X_1$ and $X_2$ do not appear independent, and that the distribution on $X_1+X_2$ does not appear purely binomial. The bound on $P[X_1+X_2=0]$ is the best available bound so far for the purpose of verifying the basis step, and gives the existence of a labeling with $s$ labels, with $s$ chosen as small as $s=\Delta(G)+1.$ It is not possible to take $s=\Delta(G),$ set $(1-P[X_1=0]) \le E[X_1]$ and get the existence of a proper edge labeling on an arbitrary graph with just $\Delta(G)$ labels. The next example uses the identical technique, but the set-up and rationale for invoking the technique is based on a different type of labeling problem. \\

\noindent Total labelings are a popular type of labeling, and there is not much progress on the constructive solution to the problem. Both the problems here are somewhat more difficult than some of the other instances of the reference set method. In both these two cases, the fibers are not strictly isolated vertices or isolated edges, and in both these cases it appears that it is necessary to use the Boolean variable instead of the standard variable; fix the value for $\beta$ in some suitable range.    \\

\noindent (Example. Total Labelings) Here, try to address the Total Labeling Conjecture. Use an auxiliary graph. The example is very unusual in that it, essentially, treats both edges and vertices as fibers. Try to use constructions to show that there are no ideas required beyond the techniques of the usual steps. \\

\noindent Step 1. First, find a regular super-graph $H_0$ of $G$ such that $\Delta(H_0)=\Delta(G).$ Then, form an auxiliary graph $H$ from $H_0,$ that is symmetric and bipartite, such that the vertices in each partite set of $H$ represent the vertices and edges of $H_0.$ Form an edge $(v,u)$ in $H$ if $(1)$ $v$ and $u$ have vertex pre-images in $H_0$ that are adjacent, $(2)$ $v$ and $u$ have vertex and edge pre-images in $H_0$ respectively and the pre-images are incident, or $(3)$ $v$ and $u$ have edge pre-images in $H_0$ that are incident. Color the edges from partite set $A$ of $H$ to partite set $B$ of $H,$ so that if $v$ is a vertex of $H_0,$ then the adjacencies of type $(1)$ are colored blue, and the adjacencies of type $(2)$ are colored red. If $v$ is an edge of $H_0$, then the edges of type $(2)$ and $(3)$ are arbitrarily colored red or blue depending on which end-point of the edge generates the adjacencies, so that all the adjacencies and incidences are colored blue for one end-point, and all the adjacencies and incidences generated by the other end-point are colored red. The partition is arbitrary for each edge. The bipartite graph $H$ need not be symmetric in the two partite sets with regard to the initial coloring. That is, the coloring is generated by $v$ in partite set $A$ for all the vertices in partite set $A$. The new graph $H$ has $n(H)=2|V(G)|+2|E(G)|.$ Establish a uniform map from the partite set $A$ of $H$ to $K_{n(H)/2}$. The vertices of partite set $A$ in $H$ are the fibers in the construction. The relationship of the comparison orbit to the physical graph is somewhat non-standard. Show that there is a proper vertex labeling in $H$ with no collisions; this will imply that there is a labeling of $G$ with no collisions. This, second, proper vertex labeling is symmetric in the partite sets of $H.$ Assume that $H$ is regular, without loss of generality. \\
 
\noindent Step 2. Use an approach analogous to the previous example. There are two identically distributed, independent random variables $X_1$ and $X_2$ at each fiber. The variable $X_1$ is generated by the blue edges, and the random variable $X_2$ is generated by the red edges. The conjunction of the random variables counts the number of duplications at the standard random variable. If the fiber has a vertex as its pre-image in $G,$ then one random variable counts vertex-edge label duplicates, and the other random variable counts vertex label duplicates in the neighborhood of the fiber's pre-image in $G.$ Use a transformation of the random variable. By selecting $s$ correctly, it follows that the value for $\beta << \Delta(G)+2.$ \\

\noindent Step 3. The basis step is the primary limitation on the problem. Use a transformation on the random variable analogous to the transformation in the Vizing Theorem example, and see that $\Delta(G)+2 \le s$ is sufficient to check the basis step. In the desired range, $\beta << \Delta(G)+2.$ The reference set hypothesis checks as $$1/|S|=[\Delta(G)+2]^{-[n(H)/2]}<e^{-1/\beta}\frac{1}{\beta^{[n(H)/2]}},$$ when $s$ is chosen $s=\Delta(G)+2.$ Both the basis step and the reference set hypothesis check, and so the example is complete, because of the construction on the graph $G$ and the graph $H,$ the auxiliary graph. \\

\newpage

\noindent \textbf{Afterword.} If there is a set of two or more graphs under the labeling or selection problem, and the edges are taken as fibers, then it may be necessary to show the collection of graphs can be packed edge-disjointly into a complete graph in order to apply the results of the reference set method. In edge-disjoint packing problems phrased as labeling problems, the cardinality of the comparison orbit is generally not well-defined, when the edges are taken as fibers, even in the case of acyclic graphs. There is a phrasing of an edge-disjoint packing problem as a labeling problem when the edges are fibers, that does give a well-defined cardinality of the comparison orbit, but it is limited to acyclic graphs. The phrasing is only available for a set of graphs that can be packed as a multigraph so that the degree sequence of the multigraph is majorized by $(n-1)^n.$ Because the phrasing of packing problems using vertices as fibers requires a separate degree condition, these phrasings generally only work for packings with two graphs, though the packing can be done inductively, using the larger maximum degree of the previously packed set of graphs at each stage of the induction. Take the example of the edge-disjoint packing problem of a star $S_{1,n-1}$ and a matching $(n/2)K_2.$ Notice that the reference set method requires a covering space as a starting point for the application of any relevant hypotheses. The value of $2 \mu =2=\beta$ in this case, and suppose $\alpha = d \alpha$. Take the edges of one of the graphs as fibers: $1/\beta << (1-2/(d \alpha)) \le P[X_1=0].$ In this case, the reference set hypothesis will be satisfied. There is no false positive result because the collection of graphs do not satisfy the hypothesis that $K_n$ covers the collection of graphs (in an edge-disjoint fashion). There are alternative hypotheses that can be used the satisfy the reference set method, and still make substantial conclusions. For instance, there is a construction available for packing sets of two or more graphs that requires an additional degree sequence condition. Also, if there are only two graphs in the edge-disjoint placement problem, there is a standard way to proceed that does not require the degree sequence condition. So, first, suppose the packing is on just two graphs, $G$ and $H,$ in the same complete graph, and take the edges of $G$ as the fibers of induction. If the condition holds that $G$ is acyclic, then proceed in a standard way to address the problem without checking the degree sequence requirement. Let the fibers of induction be the edges of $G,$ and, additionally, use the so-called $2$-body allowance, as long as $G$ and $H$ are the only elements of the packing. In this case, it appears necessary to observe an additional condition in order to phrase the problem in this fashion: For instance, when $G \cong T$ is a tree of order $k,$ the problem appears to require setting $\beta=\beta_1+\beta_2$ where $\beta_1=(1/2)2\mu/\nu_0,$ $\beta_2 = k=2{k \choose 2}/(n(T) \cdot \nu_0),$ $\nu_0=(k-1)/n(T).$  Upon adding this additional summand to the expectation, it appears that the results that follow are consistent with accepted results. Relax the condition that $G$ must be acyclic when the vertices are the fibers of induction, but again, it appears necessary to apply an additional summand to the expectation, the value $\beta_2,$ albeit with a different parameterization. Apply the $2$-body allowance, and phrase the problem using the expectation $\beta  = \Delta(G) \Delta(H)+ \beta_2,$ with the vertices as fibers, and $\beta_2 = (n-1)/2.$ Also, when the vertices are taken as the fibers of induction, it appears that the basis case has to be verified using a condition on $\Delta(G).$ Check the basis case using worst-case analysis. This analysis gives the expression $1/\beta << 1-(\kappa \Delta(G)+(n-1)/(2n))$ where $\kappa=\Delta(H)/n$ as sufficiency for the base case. Then, if there are more than two elements in the packing, see that the $2$-body allowance does not apply. To review, reconsider the same extremal example of a star and a matching that span the same complete graph. Use the $2$-body allowance. Notice that there are restrictions on the problem that are exchanged for weakening the hypotheses required of the reference set method. Reparameterize by $S_{1,2n-1} \cap nK_2.$ Measure $\mu$ in the standard fashion. Apply the $2$-body allowance, and take the edges of the star as the fibers of induction. Then, it appears necessary to apply a penalty summand to the expectation in the problem. A penalty portion applies to the expectation, otherwise there is the false positive result that these two graphs pack, and, of course, there is no edge-disjoint placement in this case. Similarly, take $V(S_{1,2n-1})$ as the fibers of induction, and apply any combination of the available sufficient conditions. The false positive result that there is an edge-disjoint placement should never be possible. The case is analogous on the matching, using the vertices or edges of the matching as the fibers of induction. There is at least one more phrasing of the edge-disjoint placement problems that can give results under the reference set method. That is, there is a standard way to proceed and get mutually edge-disjoint placements of acyclic graphs $\{H_i\} $ with $2 \le |[i]|,$ when the degree sequences do pack. Usually, when $|[i]|=2,$ the approach with the $2$-body allowance is advantageous. However, in some cases, it is necessary that $3 \le |[i]|,$ and this second phrasing of edge-disjoint placement problems will work. If the labeling problem is an edge-disjoint packing problem on $\{H_i\},$ then use the condition $\pi(\phi(\{H_i\})) \prec (n-1)^n$ is multigraphic with some of the standard conditions, including that all the graphs are acyclic, to show results. In this case, use the local topology of $\lambda K_n$ with $\lambda=1$ to cover the set of graphs. This extension of the method has some possible advantages. First, notice that there is a transformation of the random variable that is available in these situations, that permits the application of the local topology as a model. The local topology of the complete graph is a covering space, and the number of fibers doubles from the model of the complete graph. The model allows the application of $P[X_1=0]=P[X_1<2].$ Use $1/|S| \le e^{-2/\beta} [2/\beta]^{\alpha}$ with $\alpha=n(n-1)$ as the check of the reference set hypothesis, instead of the standard check where the value of $x$ is taken $x=1$ on the right-hand side. In these cases, verify the degree sequence condition $\pi(\phi(\{H_i\})) \prec (n-1)^n$ by covering some placement of the graphs with a suitable multigraph. Notice that the basis step has to be verified, and that the support on the basis step is extended to $X_1<2$. Using the model of the local topology, it does not appear viable to trim down to an exact cover of the set of graphs, and reduce $\alpha$ to $\alpha=2 \cdot \sum_{i \in [i]}~m_i$. That is, see \cite{JG3}, previous version; the contrary assertion may not be correct. Rather, the advantage in sparse packings is that $\beta<<\sqrt{n-1}.$ The reference set method appears to have a number of applications in graph labeling problems when the domain of the labeling function, $\phi,$ is the vertex set $V(G)$ of the graph $G,$ or when the domain of $\phi$ is $E(G).$

\end{document}